\documentclass[10pt,conference]{IEEEtran}
\IEEEoverridecommandlockouts

\usepackage{cite}
\usepackage{amsmath,amssymb,amsfonts,mathtools,braket}
\usepackage{hyperref}
\usepackage{algorithm,algorithmic}
\usepackage{graphicx}
\usepackage{textcomp}
\usepackage{xcolor}
\usepackage[normalem]{ulem}

\usepackage{subcaption}
\usepackage{float}

\DeclareMathOperator*{\argmin}{arg\,min}

\def\BibTeX{{\rm B\kern-.05em{\sc i\kern-.025em b}\kern-.08em
    T\kern-.1667em\lower.7ex\hbox{E}\kern-.125emX}}
\begin{document}

\title{An Adaptive Weighted QITE-VQE Algorithm for Combinatorial Optimization Problems}

\author{
\IEEEauthorblockN{
Ningyi Xie\IEEEauthorrefmark{1},
Xinwei Lee\IEEEauthorrefmark{2},
Tiejin Chen\IEEEauthorrefmark{3},
Yoshiyuki Saito\IEEEauthorrefmark{4},
Nobuyoshi Asai\IEEEauthorrefmark{5},
Dongsheng Cai\IEEEauthorrefmark{6}
}

\IEEEauthorblockA{\IEEEauthorrefmark{1}\textit{Graduate School of Science and Technology}, \textit{University of Tsukuba}, Ibaraki, Japan}

\IEEEauthorblockA{\IEEEauthorrefmark{2}\textit{School of Computing and Information Systems}, \textit{Singapore Management University}, Singapore, Singapore}

\IEEEauthorblockA{\IEEEauthorrefmark{3}\textit{School of Computing and Augmented Intelligence}, \textit{Arizona State University}, Tempe, USA}

\IEEEauthorblockA{\IEEEauthorrefmark{4}ARGO GRAPHICS Inc., Tokyo, Japan}

\IEEEauthorblockA{\IEEEauthorrefmark{5}\textit{School of Computer Science and Engineering}, \textit{University of Aizu}, Fukushima, Japan}

\IEEEauthorblockA{\IEEEauthorrefmark{6}\textit{Faculty of Engineering, Information and Systems}, \textit{University of Tsukuba}, Ibaraki, Japan}

\thanks{\IEEEauthorrefmark{1} \href{mailto:nyxie@cavelab.cs.tsukuba.ac.jp}{nyxie@cavelab.cs.tsukuba.ac.jp}}

\thanks{\IEEEauthorrefmark{6} Currently affiliated with Nagoya University of Commerce and Business}

}

\maketitle

\begin{abstract}
    The variational quantum eigensolver (VQE) is an algorithm for finding the ground states of a given Hamiltonian. Its application to binary-formulated combinatorial optimization (CO) has been widely studied in recent years. However, typical VQE approaches for CO problems often suffer from local minima or barren plateaus, limiting their ability to achieve optimal solutions. The quantum imaginary time evolution (QITE) offers an alternative approach for effective ground-state preparation but requires large circuits to approximate non-unitary operations. Although compressed QITE (cQITE) reduces circuit depth, accumulated errors eventually cause energy increases. To address these challenges, we propose an Adaptive Weighted QITE-VQE (AWQV) algorithm that integrates the VQE gradients with the cQITE updates through an adaptive weighting scheme during optimization. In numerical simulations for MaxCut on unweighted regular graphs, AWQV achieves near-optimal approximation ratios, while for weighted Erdős-Rényi instances, it outperforms the classical Goemans-Williamson algorithm.
\end{abstract}


\section{Introduction}\label{sec:introduction}
Quantum computing as a novel computational paradigm leverages superposition and entanglement, potentially offering significant speedups compared to classical counterparts \cite{montanaro2016quantum,grover1996fast,shor1999polynomial}. However, the field is currently in the noisy intermediate-scale quantum (NISQ) era \cite{preskill2018quantum}, facing significant constraints, including limited qubit counts and noise processes that restrict circuit depth. In this context, variational quantum eigensolver (VQE) \cite{peruzzo2014variational} has emerged as a leading candidate for demonstrating quantum advantage due to its flexibility in quantum resources, allowing adjustable circuit depth.

VQE approximates the ground state of a given Hamiltonian by assuming an ansatz with trainable parametrized unitary operations and utilizing classical optimizers to tune these parameters toward energy minimization. An extensively explored application of VQE is combinatorial optimization (CO), where the given CO problem is binary formulated, allowing a mapping to the computational basis Hamiltonian. Farhi et al. \cite{farhi2014quantum} proposed the quantum approximate optimization algorithm (QAOA) for CO problems, which shares a similar variational framework with VQE but employs a problem-dependent ansatz inspired by quantum adiabatic computing \cite{farhi2000quantum}. Following this, a plethora of QAOA variants have been proposed to effectively address constrained CO problems \cite{hadfield2019quantum,wang2020xy,bartschi2020grover,herman2023constrained,xie2024feasibility}, mitigate or leverage device noise for improved performance \cite{sack2024large,maciejewski2024improving}, and reduce quantum resource requirements \cite{tan2021qubit,Glos2022,herrman2022multi,vijendran2024expressive,wang2025imaginary}. Notably, the expressive QAOA (XQAOA) proposed by Vijendran et al. \cite{vijendran2024expressive} reports a superior performance on maximum cut (MaxCut) problems of regular graphs with degrees greater than 4, surpassing the classical state-of-the-art approximation algorithm, the Goemans-Williamson (GW) algorithm \cite{goemans1995improved}. Beyond QAOA-like ansatze, Wang et al. \cite{wang2025imaginary} present a quantum imaginary time evolution (QITE)-inspired circuit for MaxCut problems, named imaginary Hamiltonian variational ansatz (iHVA), which also shows better results than the GW algorithm on regular graphs with degrees fewer than 5. However, VQEs for CO problems typically face optimization challenges, including local minima and barren plateaus in the parameter landscape \cite{mcclean2018barren,lee2021parameters,blekos2024review}. The successes reported in \cite{vijendran2024expressive} and \cite{wang2025imaginary} rely not only on the reasonable ansatz designs but also on extensive parameter optimization with multiple random starting points. More broadly, parameter initialization and optimization strategies remain an active area of research for VQE implementations \cite{zhou2020quantum,alam2020accelerating,lee2021parameters,moussa2022unsupervised,xie2023quantum,lee2023depth}.

Unlike variational methods prone to local minima, QITE offers a principled alternative with a guarantee of convergence to the ground state \cite{motta2020determining}. Ideally, the QITE drives the state $\ket{\psi_\tau} = c^{-\frac{1}{2}}e^{-\tau H}\ket{\psi_0}$ to approach the ground state of Hamiltonian $H$ as $\tau\rightarrow \infty$, given that the initial state $\ket{\psi_0}$ has a nonzero overlap with the ground state. Here, $e^{-\tau H}$ represents a non-unitary operator, and $c^{-\frac{1}{2}}$ serves as the normalization factor. Motta et al. \cite{motta2020determining} propose an algorithm to approximate a non-unitary operation using a sequence of unitary operations implementable on quantum devices. However, the algorithm divides the evolution time $\tau$ into $N$ small time steps such that $\tau = N\Delta\tau$, requiring $N$ sequential unitary operations to mimic $c^{-\frac{1}{2}}e^{-\tau H}$, resulting in a deep circuit unsuitable for NISQ devices. Variational QITE (VarQITE) \cite{McArdle2019} employs McLachlan’s variational principle \cite{mclachlan1964variational,broeckhove1988equivalence} to update circuit parameters for approximating imaginary time evolution, thereby reducing circuit depth. Similarly, the quantum natural gradient (QNG) \cite{stokes2020quantum} method bridges the gap between VQE and imaginary time evolution by aligning parameter updates with the quantum Fisher information metric, and QNG is equivalent to VarQITE for certain ansatze. However, the applicability of VarQITE to CO problems and its comparative performance against QITE in ground state identification remain open questions and merit further investigation. Nishi et al. \cite{nishi2021implementation} and Gomes et al. \cite{gomes2020efficient} propose similar approaches that compress QITE steps into a fixed-depth circuit using reverse Suzuki-Trotter decomposition, known as compressed QITE (cQITE) or step-merged QITE (smQITE), in which each step naturally corresponds to a parameter update. While cQITE significantly reduces circuit depth, compression errors accumulate at each time step, eventually causing energy increases that prevent convergence to the ground state. A straightforward approach for further convergence is to apply gradient-based optimization once energy increases, using cQITE to provide the initial parameters for VQE. However, when the initialization from cQITE is not sufficiently strong, the subsequent VQE optimization may still become trapped in local minima, as shown in Section~\ref{sec:unweighted} and Appendix~\ref{app:qiv}. 

In this work, we propose an \textbf{A}daptive \textbf{W}eighted \textbf{Q}ITE-\textbf{V}QE (AWQV) algorithm, aiming to use cQITE-derived updates to continuously guide VQE toward regions near the global minimum and maintain convergence as compression errors in cQITE accumulate. 
The core of AWQV lies in an adaptive weighting scheme that modifies the parameter gradient by incorporating the cQITE update direction. 
Initially, the scheme favors the cQITE component to provide an initialization. As the energy converges, the weight is gradually shifted toward the gradient component, thereby mitigating the impact of accumulated errors introduced by the compression, facilitating convergence. We apply the AWQV to the MaxCut problems. On unweighted regular graphs, AWQV consistently achieves near-optimal approximation ratios in expectation based on the best solution obtained from 10 samples drawn from the prepared state. In contrast, VQEs with gradient-based optimizers converge to excited states in several instances, resulting in a near-zero ground state probability.   Furthermore, we compare AWQV with the GW algorithm on weighted Erdős-Rényi graphs, where AWQV demonstrates lower failure rates (20 versus 29 failures out of 432 instances) in yielding optimal solutions under the same sampling budget.

The rest of this paper is structured as follows: Section 2 gives background on MaxCut, VQE, and QITE. Section 3 describes the proposed AWQV algorithm. Section 4 reports the simulation results, and Section 5 concludes the paper.

\section{Preliminaries}\label{sec:preliminaries}
\subsection{MaxCut}
The MaxCut problem has been widely adopted as a benchmark for evaluating quantum optimization algorithms \cite{zhou2020quantum,moussa2022unsupervised,xie2023quantum,lee2021parameters,lee2023depth,alam2020accelerating}. The objective of the MaxCut problem is to find a partition of the vertices of a given graph into two sets such that the total weight of the edges crossing between the two sets is maximized. Given a graph $G = (V, E)$, where $V$ and $E$ denote the sets of vertices and edges respectively, the problem can equivalently be formulated as 
\begin{equation}
    \argmin_{\mathbf{x} \in \{0,1\}^{|V|}} C(\mathbf{x}),\;C(\mathbf{x}) = -\sum_{(i,j)\in E} w_{ij} (x_i - x_j)^2,
\end{equation}
where $w_{ij}$ represents the weight of the edge $(i,j)$. In the case of unweighted graphs, $w_{ij} = 1$ for all $(i,j)\in E$. By replacing each binary variable $x_i$ with $(I-Z_i )/2$, where $Z_i$ denotes the Pauli-$Z$ operation acting on the $i$-th qubit, the cost function can be mapped to a Hamiltonian $H$ such that $\braket{\mathbf{x}|H|\mathbf{x}} = C(\mathbf{x})$, for all computational basis states $\ket{\mathbf{x}}$.

\subsection{Variational quantum eigensolver}
The VQE approximates the ground state of a Hamiltonian by preparing a parameterized quantum state $ \ket{\psi(\boldsymbol{\theta})} = U(\boldsymbol{\theta})\ket{\psi_0}$, where $\ket{\psi_0}$ is an easy-to-prepare initial state and $U(\boldsymbol{\theta})$ is a sequence of parameterized unitary operations with tunable parameters $\boldsymbol{\theta}$. The choice of $U(\boldsymbol{\theta})$ is generally guided by the structure of the target Hamiltonian \cite{farhi2014quantum,wang2025imaginary} or designed to be hardware-efficient \cite{zhang2022variational,yan2024light}. Given a Hamiltonian $H$, the optimization objective is to
\begin{equation}
    \argmin_{\boldsymbol{\theta}} \braket{\psi(\boldsymbol{\theta})|H|\psi(\boldsymbol{\theta})}.
\end{equation}
Generally, VQE employs a classical optimizer to perform this minimization. For example, using the gradient descent (GD) \cite{ruder2016overview} algorithm, the parameters at the $s$-th iteration are updated as
\begin{equation}
    \boldsymbol{\theta}^{(s)} = \boldsymbol{\theta}^{(s-1)} - \eta \nabla_{\boldsymbol{\theta}^{(s-1)}} ,
\end{equation}
where $\eta$ denotes the learning rate and the gradient $\nabla_{\boldsymbol{\theta}} \coloneqq \nabla_{\boldsymbol{\theta}} \braket{\psi(\boldsymbol{\theta})|H|\psi(\boldsymbol{\theta})}$ can be estimated using the parameter-shift rule \cite{mitarai2018quantum,schuld2019evaluating}.

\subsection{Quantum imaginary time evolution}
Given a Hamiltonian $H = \sum_{l=1}^{L} H_l $, which represents a sum of $L$ weighted Pauli strings. The original Trotterized QITE method \cite{motta2020determining} applies the Suzuki-Trotter decomposition to the imaginary time evolution operator $e^{-\tau H}$, resulting in
\begin{equation}
    e^{-\tau H} = \left( e^{-\Delta\tau H_1} e^{-\Delta\tau H_2} \cdots \right)^N + \mathcal{O}(\Delta \tau), \quad \tau = N\Delta\tau.
\end{equation}
Then, each non-unitary evolution $e^{-\Delta\tau H_l}$ is approximated using a sequence of unitary operations. For CO problems, where the given Hamiltonian $H$ is diagonal in the computational basis, all terms commute with each other, and thus no Trotter errors are introduced during the decomposition. Following \cite{alam2023solving,bauer2024combinatorial}, we directly approximate $e^{-\Delta\tau H}$ without further decomposition into subterms. 

\begin{figure}[t]
    \includegraphics[width=0.85\linewidth]{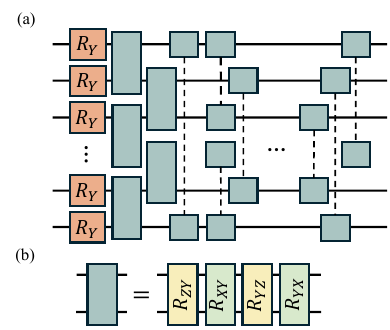}
    \caption{(a) The overall circuit structure of P2A, consisting of single-qubit $R_Y$ rotations on each qubit and two-qubit entanglement blocks acting on every qubit pair, with each gate independently parameterized. Using the round-robin scheduling \cite{rasmussen2008round}, the circuit depth scales as $\mathcal{O}(n)$, where $n$ is the number of qubits. (b) Entanglement block of P2A, composed of $R_{ZY}$ and $R_{XY}$ rotations.}
    \label{fig:circuits}
\end{figure}

Let $\ket{\psi_{s-1}}$ denote the state obtained after evolving from the initial state \(\ket{\psi_0}\) for $s-1$ imaginary time steps of size $\Delta\tau$. The target at the $s$-th step is given by
\begin{equation}
     \argmin_{\boldsymbol{\phi}^{(s)}} \left\lVert \frac{e^{-\Delta\tau H}\ket{\psi_{s-1}}}{\sqrt{\braket{\psi_{s-1}|e^{-2\Delta\tau H}|\psi_{s-1}}}}-e^{-i\Delta\tau A(\boldsymbol{\phi}^{(s)})}\ket{\psi_{s-1}}\right\rVert^2,
\end{equation}
where 
\begin{equation}
    A(\boldsymbol{\phi}^{(s)}) = \sum_{P \in \mathcal{P}} \phi^{(s)}_P P
\end{equation}
represents a weighted sum of Pauli strings from a predefined set $\mathcal{P}$. \cite{motta2020determining} shifts this problem to a linear system, $S^{(s)} \boldsymbol{\phi}^{(s)} = \mathbf{b}^{(s)}$, where 
\begin{equation}
    \begin{aligned}
        S_{PP^{\prime}}^{(s)} & = \text{Re}\left(\braket{\psi_{s-1}|P^{\dagger}P^{\prime}|\psi_{s-1}}\right),\\
        b^{(s)}_{P} & = \text{Im}\left(\braket{\psi_{s-1}|P^{\dagger}H|\psi_{s-1}}\right).
    \end{aligned}\label{eqn:linear_system}
\end{equation}
Let $\hat{\boldsymbol{\phi}}^{(s)} =  \argmin_{\boldsymbol{\phi}^{(s)}}  \left\lVert S^{(s)} \boldsymbol{\phi}^{(s)} - \mathbf{b}^{(s)}\right\rVert^2$, 
decompose $e^{-i\Delta\tau A(\hat{\boldsymbol{\phi}}^{(s)})}$ as
\begin{equation}
    e^{-i\Delta\tau A(\hat{\boldsymbol{\phi}}^{(s)})} = \prod_{P \in \mathcal{P}}e^{-i\Delta\tau \hat{\phi}^{(s)}_P P} + \mathcal{O}(\Delta\tau^2),
\end{equation} the state after $s$ steps is 
\begin{equation}
    \ket{\psi_{s}} = \prod_{P \in \mathcal{P}}e^{-i\Delta\tau \hat{\phi}^{(s)}_P P}\ket{\psi_{s-1}}.
\end{equation}

The cQITE modifies the original QITE procedure by composing the unitary evolution $e^{-i\Delta\tau A(\phi)}$ onto the merged operations at each time step. Accordingly, the state after s steps is
\begin{equation}
    \ket{\psi_s} = \prod_{P \in \mathcal{P}}e^{-i\Delta\tau \sum_{m=1}^{s}\hat{\phi}^{(m)}_P P}\ket{\psi_{0}}.
\end{equation}
As the circuit is fixed, cQITE can also be viewed as a variational method with a parametrized unitary operator, 
\begin{equation}
    U(\boldsymbol{\theta}) = \prod_{P \in \mathcal{P}}e^{-i \frac{\theta_P}{2} P},\label{eqn:param_unitary}
\end{equation}
and the parameters at the $s$-th iteration are updated as
\begin{equation}
    \boldsymbol{\theta}^{(s)} = \boldsymbol{\theta}^{(s-1)} + 2\Delta\tau \boldsymbol{\phi}^{(s)},\; \boldsymbol{\theta}^{(0)} = \mathbf{0} \label{eqn:cqite_update}
\end{equation}
with each imaginary time step corresponding to one update, where $2\Delta\tau$ is analogous to a learning rate.

\begin{figure*}[t]
    \centering
    \includegraphics[width=0.7\linewidth]{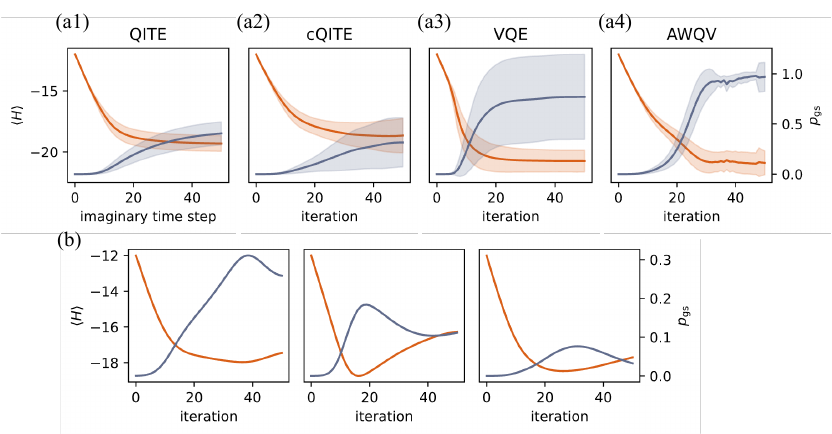}
    \caption{(a) Energy expectation value $\braket{H} \coloneqq \braket{\psi | H | \psi} $ (orange) and ground-state probability $p_{\text{gs}}$ (blue) for QITE, cQITE, VQE, and AWQV using the P2A initialized from $\ket{+}^{\otimes 16}$ on MaxCut Hamiltonians generated from 48 random 16-vertex 3-regular graphs. Solid lines represent the mean values at each step, and shaded areas indicate the standard deviations. $\Delta\tau$ is set to $0.05$ for QITE and cQITE. VQE employs the GD optimizer. (b) Curves of cQITE on three instances.}
    \label{fig:curve}
\end{figure*}

The set of Pauli strings $\mathcal{P}$ determines the evolution operations in QITE and the parameterized unitary in cQITE, respectively. \cite{nishi2021implementation} presents a construction of $\mathcal{P}$ for nonlocal approximation as
\begin{equation}
    \mathcal{P}_D \!=\!\left\{ \!P\!=\!P_{q_1}P_{q_2}\cdots P_{q_D}\middle\vert \mathbf{q}\!\in\!\binom{[n]}{D} ,\, P \!\in\! \{I,X,Y,Z\}^D\! \right\},
\end{equation}
where $n$ denotes the number of qubits, and $D$ is the domain size that limits the number of qubits each Pauli string acts on. Here, $\binom{[n]}{D}$ represents the set of all possible combinations of $D$ elements chosen from $[n]=\{1,2,\cdots,n\}$, where each combination is a vector with entries in ascending order  (e.g., $\binom{[3]}{2} = \{[1,2],[1,3],[2,3]\}$). Since the operator $e^{-\Delta\tau H}$ is not further decomposed into local qubit operations in this setting, the local-Hamiltonian-independent construction of $\mathcal{P}_D$ is well-suited for CO problems. For computational basis Hamiltonian, $b_P = 0$ unless $P$ contains an odd number of Pauli-$Y$ \cite{motta2020determining}. Hence, we define $\mathcal{P}_D^\ast$ as the reduced version of $\mathcal{P}_D$, obtained by eliminating all Pauli strings with an even number of $Y$ operators, and term the circuit determined by it as the $\mathcal{P}_D^\ast$ ansatz (P$D$A), with Fig.~\ref{fig:circuits} illustrating P2A for $D=2$. Specifically, \cite{alam2023solving} applies P1A (named the linear ansatz in \cite{alam2023solving}) to MaxCut problems. Since the operations determined by $\mathcal{P}_1^\ast$ are exclusively $R_Y$ rotation gates that commute with each other, QITE with P1A naturally behaves as cQITE.

Figure~\ref{fig:curve}(a1)-(a3) shows the progress of the energy and ground-state probability, defined as
\begin{equation}
    p_{\text{gs}} = \sum_{\mathbf{x}^\ast} \left\vert \braket{\mathbf{x}^\ast|\psi} \right\vert^2,\; \mathbf{x}^\ast = \argmin_{\mathbf{x}}C(\mathbf{x}), \label{eqn:pgs}
\end{equation}
during the imaginary time evolution or optimization of QITE, cQITE, and VQE with P2A for MaxCut Hamiltonians, where $\ket{\psi}$ denotes an intermediate state. As the number of iterations or imaginary time steps increases, all three methods exhibit a general trend of decreasing energy and increasing $p_{\text{gs}}$, with VQE achieving the fastest energy reduction. However, the standard deviation of VQE's $p_{\text{gs}}$ exhibits a sharp increase after approximately 5 iterations. This suggests inconsistent performance across different VQE instances, with some achieving high ground state probabilities while others becoming trapped in local minima at an early stage. In contrast, cQITE initially mirrors QITE's behavior but gradually exhibits greater fluctuations in energy and $p_{\text{gs}}$. Figure~\ref{fig:curve}(b) plots the curves of cQITE on three instances, where the energy increases and $p_{\text{gs}}$ decreases after certain iterations, indicating that cQITE lacks the ability to converge to the ground state. 

Motivated by the early-stage effectiveness of cQITE and the convergence tendency of VQE, we propose an adaptive weighting scheme. As shown in Fig.~\ref{fig:curve}(a4), compared to cQITE and VQE, the proposed AWQV exhibits lower variance in both energy and $p_{\text{gs}}$ across different instances, indicating more consistent performance.

\section{Adaptive weighted QITE-VQE}\label{sec:awqv}
The proposed AWQV algorithm builds upon VQE with gradient-based optimization. Like VQE, AWQV prepares a parameterized state $\ket{\psi(\boldsymbol{\theta})} = U(\boldsymbol{\theta})\ket{\psi_0}$, but with $U(\boldsymbol{\theta})$ specifically constructed from a set of Pauli strings as Eq. (\ref{eqn:param_unitary}) and $\ket{\psi_0}$ overlaps with the ground state of the target Hamiltonian $H$. The parameter update mechanism in AWQV combines the energy gradient and the cQITE update term in a weighted manner. At the $s$-th iteration, the parameters are updated according to
\begin{equation}
    \boldsymbol{\theta}^{(s)} = \boldsymbol{\theta}^{(s-1)} - \eta\tilde{\nabla}_{\boldsymbol{\theta}^{(s-1)}},\; \boldsymbol{\theta}^{(0)} = \boldsymbol{0}
\end{equation}
where $\eta$ denotes the learning rate, and $\tilde{\nabla}_{\boldsymbol{\theta}^{(s-1)}}$ denotes the modified update direction, given by
\begin{equation}
    \tilde{\nabla}_{\boldsymbol{\theta}^{(s-1)}}  = w(s)\frac{\|\hat{\boldsymbol{\phi}}^{(s)}\|}{\|\nabla_{\boldsymbol{\theta}^{(s-1)}}\|}\nabla_{\boldsymbol{\theta}^{(s-1)}}-2(1-w(s))\hat{\boldsymbol{\phi}}^{(s)}.\label{eqn:awqv_update}
\end{equation}
    At each iteration, the cQITE update $\hat{\boldsymbol{\phi}}^{(s)}$ is obtained by solving the linear system in Eq.~\ref{eqn:linear_system}, given $\ket{\psi_{s-1}} = \ket{\psi(\boldsymbol{\theta}^{(s-1)})}$. The factor, $\frac{\|\hat{\boldsymbol{\phi}}^{(s)}\|}{\|\nabla_{\boldsymbol{\theta}^{(s-1)}}\|}$, is introduced to align the scales of the gradient and the cQITE update. $w(s)$ is a weighting function, set to $0$ at the first step with $w(1) = 0$, enabling the cQITE update to initialize the parameters of $U$. For $s>1$ iteration, $w(s)$ is given by
\begin{equation}
    w(s) = \mu w(s-1) + (1-\mu)\left(1 - \frac{\lambda \delta(s-1)}{\frac{1}{s-1}\sum_{l=1}^{s-1}\delta(l)}\right) \label{eqn:w_update}
\end{equation}
where 
\begin{equation}
    \begin{aligned}
        \delta(l) & = \braket{H}^{(l-1)} - \braket{H}^{(l)},\\
        \braket{H}^{(l)} & = \braket{\psi(\boldsymbol{\theta}^{(l)})|H|\psi(\boldsymbol{\theta}^{(l)})}. \label{eqn:delta_l}
    \end{aligned}
\end{equation}
We further restrict $w(s)$ as a monotonically non-decreasing function ranging in $[0,1]$ by
\begin{equation}
    w(s) \leftarrow \min(\max(w(s),w(s-1)),1). \label{eqn:min_max}
\end{equation}
The hyperparameter, $\mu \in (0,1)$, determines the contribution of the previous weight $w(s-1)$ to the current weight $w(s)$. A larger $\mu$ results in a slower change in $w(s)$, promoting smoother and more stable weight transitions across iterations. Another hyperparameter, $\lambda > 0$, conditions the increasing of weight, as
\begin{equation}
    w(s)>w(s-1),\text{ if }\frac{\delta(s-1)}{\frac{1}{s-1}\sum_{l=1}^{s-1}\delta(l)} < \frac{1-w(s-1)}{\lambda}. \label{eqn:w_increase}
\end{equation}
When the energy convergence becomes slower, $\frac{\delta(s-1)}{\frac{1}{s-1}\sum_{l=1}^{s-1}\delta(l)}$ decreases, and the weight is shifted toward the gradient component. This behavior also reduces the threshold, $\frac{1-w(s-1)}{\lambda}$, meaning that the subsequent energy decrease needs to be even slower in order to further shift the weight, ensures that cQITE continues to guide the search until the energy finally converges.

We use the parameters that achieved the lowest energy during the optimization to prepare the quantum state for sampling $M$ times, then select the solution with the best objective value.  The detailed procedure of AWQV for solving CO problems is summarized in Algorithm~\ref{alg:awqv}.

\begin{algorithm}[t]
    \caption{AWQV for CO Problems}\label{alg:awqv}
    \begin{algorithmic}[1]
    \REQUIRE Objective function $C$, ansatz $U(\boldsymbol{\theta})$, initial state $\ket{\psi_0}$, learning rate $\eta$, weighting function $w$, maximum iterations $N$, number of samples $M$
    \ENSURE Solution $\mathbf{x}$
    
    \STATE Formulate the Hamiltonian $H$ from $C$
    \STATE Initialize $\boldsymbol{\theta}^{(0)} \gets \boldsymbol{0}$
    \STATE Set $E_{\mathrm{best}} \gets \braket{\psi(\boldsymbol{\theta}^{(0)})|H|\psi(\boldsymbol{\theta}^{(0)})}$, $\boldsymbol{\theta}_{\mathrm{best}} \gets \boldsymbol{\theta}^{(0)}$
    
    \FOR{$s = 1$ to $N$}
        \STATE Compute $\nabla_{\boldsymbol{\theta}^{(s-1)}}$ via parameter-shift rule
        \STATE Solve the linear system in Eq.~(\ref{eqn:linear_system}) to obtain $\hat{\boldsymbol{\phi}}^{(s)}$
        \STATE Update parameters along the gradient direction: \\ $\boldsymbol{\theta}^{(s)} \gets \boldsymbol{\theta}^{(s-1)} - \eta w(s) \frac{\|\hat{\boldsymbol{\phi}}^{(s)}\|}{\|\nabla_{\boldsymbol{\theta}^{(s-1)}}\|} \nabla_{\boldsymbol{\theta}^{(s-1)}}$
        \STATE Update parameters along the cQITE direction: \\ $\boldsymbol{\theta}^{(s)} \gets \boldsymbol{\theta}^{(s)} + 2 \eta (1-w(s)) \hat{\boldsymbol{\phi}}^{(s)}$
        \STATE Estimate energy $E^{(s)} = \braket{\psi(\boldsymbol{\theta}^{(s)})|H|\psi(\boldsymbol{\theta}^{(s)})}$
        \IF{$E^{(s)} < E_{\mathrm{best}}$}
            \STATE Update $E_{\mathrm{best}} \gets E^{(s)}$, $\boldsymbol{\theta}_{\mathrm{best}} \gets \boldsymbol{\theta}^{(s)}$
        \ENDIF
    \ENDFOR
    \STATE Sample $M$ bitstrings $\{\mathbf{x}^{(1)}, \ldots, \mathbf{x}^{(M)}\}$ by measuring $\ket{\psi(\boldsymbol{\theta}_{\mathrm{best}})}$ in the computational basis
    \STATE Select $\mathbf{x} \gets \mathbf{x}^{m^\ast}$, $m^{\ast} = \argmin_{m \in \{1, \ldots, M\}} C(\mathbf{x}^{(m)})$
    \RETURN $\mathbf{x}$
    \end{algorithmic}
\end{algorithm}

\begin{figure*}[t]
    \centering
    \includegraphics[width=0.95\linewidth]{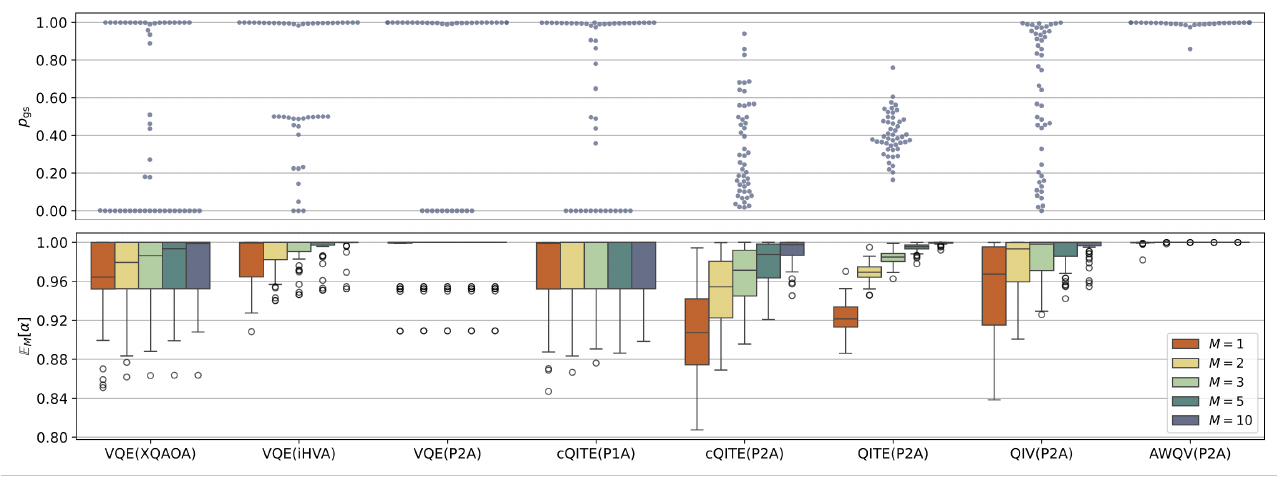}
    \caption{$p_{\text{gs}}$ (upper) and $\mathbb{E}_M[\alpha]$ with $M=1,2,3,5,10$ (lower) of states prepared by VQE, cQITE, QITE, QIV, and the proposed AWQV for MaxCut on 48 randomly generated 16-vertex 3-regular graphs. In the $p_{\text{gs}}$ plot, each point corresponds to a single instance. The $\mathbb{E}_M[\alpha]$ results are summarized in boxplots.}
    \label{fig:n16reg3}
\end{figure*}

\section{Simulation Results}\label{sec:experiments}
In this section, we evaluate the performance of the proposed AWQV algorithm on MaxCut problems using a statevector simulator \cite{Zhang2023tensorcircuit}, and compare its results with those of existing quantum approaches and the classical GW algorithm.

\subsection{Evaluation metrics}
The state prepared by the proposed quantum approaches is evaluated based on two metrics:  
\begin{itemize}
    \item Approximation ratio, $\alpha$. Given a minimization objective function $C$, let the distinct objective values over the search space $\mathcal{X}$ be sorted as
    \begin{equation}
        \begin{aligned}
            & C_1 < C_2 < \cdots < C_K,\\
            & \{C_1,C_2,\cdots,C_K\} = \left\{C(\mathbf{x})\middle\vert \mathbf{x} \in \mathcal{X}\right\}
        \end{aligned}
    \end{equation}
    where $K$ denotes the number of distinct objective values. For a solution $\mathbf{x}$, the approximation ratio $\alpha$ is given by
    \begin{equation}
        \alpha = \frac{C(\mathbf{x}) - C_K}{C_1 - C_K}.
    \end{equation}
    An $\alpha$ value closer to $1$ indicates a better approximation to the optimum.
    In the context of quantum approaches, we sample $M$ times from the prepared state $\ket{\psi}$ and select the best solution. The expected value of the highest approximation ratio $\alpha$ among the $M$ samples is used to evaluate the quality of $\ket{\psi}$, and is defined as
    \begin{equation}
        \mathbb{E}_M[\alpha] = \frac{\sum_{k=1}^{K}C_k\left[P_{\geq}(C_k)^M-P_{>}(C_k)^M\right]-C_K}{C_1-C_K},
    \end{equation}
    where
    \begin{equation}
        \begin{aligned}
            P_{\geq}(C_k) &= \sum_{\mathbf{x}:C(\mathbf{x})\geq C_k}\lvert\braket{\mathbf{x}|\psi}\rvert^2,\\
            P_{>}(C_k) &= \sum_{\mathbf{x}:C(\mathbf{x})>C_k}\lvert\braket{\mathbf{x}|\psi}\rvert^2.
        \end{aligned}
    \end{equation}
    \item Ground state probability, $p_{\text{gs}}$. As defined in Eq.~(\ref{eqn:pgs}), \(p_{\text{gs}}\) represents the probability of sampling the ground state from the prepared quantum state, determining the expected number of samples required to obtain the optimal solution, which is $\frac{1}{p_{\text{gs}}}$.
\end{itemize}

\begin{figure}[t]
    \centering
    \includegraphics[width=0.95\linewidth]{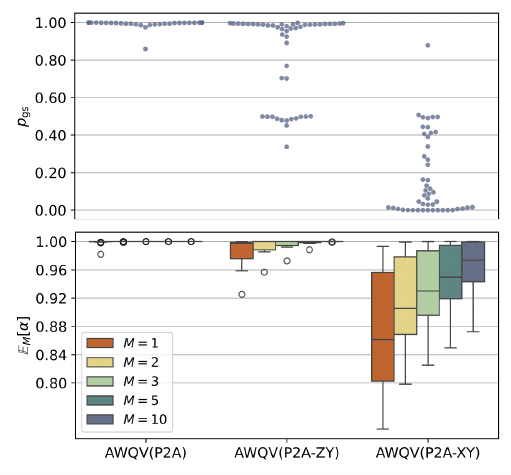}
    \caption{$p_{\text{gs}}$ (upper) and $\mathbb{E}_M[\alpha]$ with $M=1,2,3,5,10$ (lower) of states prepared by AWQV(P2A), AWQV(P2A-ZY) and AWQV(P2A-XY) for MaxCut on 48 randomly generated 16-vertex 3-regular graphs.}
    \label{fig:n16reg3_xyzy}
\end{figure}

\begin{figure*}[t]
    \centering
    \includegraphics[width=0.95\linewidth]{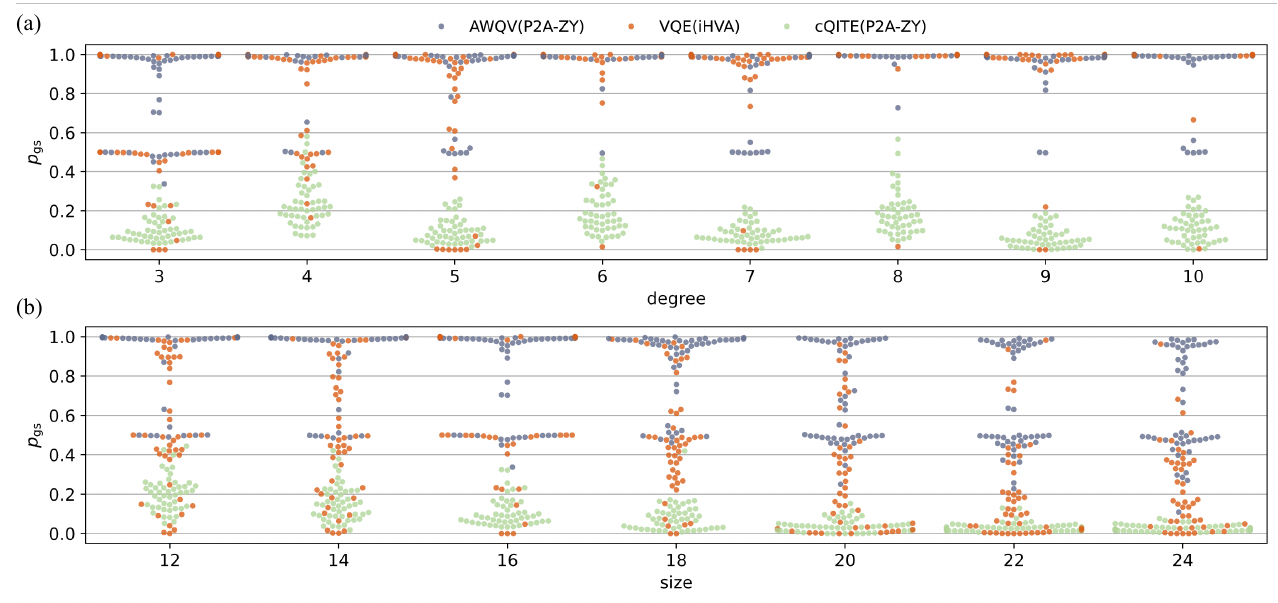}
    \caption{$p_{\text{gs}}$ values of states prepared by AWQV(P2A-ZY) cQITE(P2A-ZY) and VQE(iHVA) for MaxCut: (a) 16-vertex regular graphs with degrees ranging from 3 to 10, and (b) 3-regular graphs with sizes ranging from 12 to 24 vertices. Each type of graph consists of 48 randomly generated instances.}
    \label{fig:degree_size}
\end{figure*}

\subsection{MaxCut on unweighted graphs}\label{sec:unweighted}
We first benchmark AWQV against existing quantum algorithms, including QITE, cQITE, and VQE with GD optimizer, on 48 randomly generated 16-vertex 3-regular graphs to show its enhanced ground state convergence capabilities. The P2A ansatz is primarily adopted for all methods; in addition, we include comparisons with VQE using a 1-layer XQAOA ansatz employing an ``X=Y'' mixer \cite{vijendran2024expressive}, VQE using a 1-layer iHVA ansatz \cite{wang2025imaginary}, and cQITE using the P1A ansatz \cite{alam2023solving}. Furthermore, we introduce a simple variant of AWQV, referred to as QITE-initialized VQE (QIV), where the weighting function $w(s)$ is treated as a binary step function: once the energy begins to increase under cQITE updates, the update switches to follow the gradient direction, i.e., using the best parameters obtained from cQITE as initialization. For clarity, we denote each method together with its corresponding ansatz, such as AWQV(P2A). Regarding the initial states, all methods employ the $\ket{+}^{\otimes n}$ as the initial state, except for cQITE(P1A), which uses $\ket{0}\otimes\ket{+}^{\otimes (n\!-\!1)}$ following the setup in \cite{alam2023solving}. Here, $n$ denotes the number of qubits. The parameters of VQE(P2A) and VQE(iHVA) are initialized using a single step of QITE, while VQE(XQAOA) initializes its parameters by sampling from a standard normal distribution, as the $R_{ZZ}$ rotations in XQAOA correspond to Pauli strings without Pauli-$Y$. Table~\ref{tab:train} summarizes the optimization and simulation settings used in this study. For AWQV, the weighting function is configured with $\mu = 0.9$ and $\lambda = 1$.

Figure~\ref{fig:n16reg3} shows the $p_{\text{gs}}$ and $\mathbb{E}_M[\alpha]$ results. AWQV achieves $p_{\text{gs}}$ values close to 1 across all instances, except for one instance where $p_{\text{gs}}\! =\! 0.859$. With one additional sampling (i.e., $M\!=\!2$), AWQV is expected to quasi-guarantee the optimal solution. In contrast, the $p_{\text{gs}}$ values of VQE(XQAOA) and VQE(P2A) are concentrated near 1 or 0, suggesting that some prepared states converge to excited states. In particular, VQE(XQAOA) performs the worst on this problem set, with only nearly half of the instances achieving the optimal solution in expectation as the number of samples increases. cQITE behaves similarly to the VQE methods, and we prove its equivalence to VQE with the GD optimizer in Appendix~\ref{app:proof}. VQE(iHVA) performs the best among gradient-based methods, with most $p_{\text{gs}}$ values distributed around 1 and 0.5. However, there still exist three instances stuck at 0, appearing as outliers in the $\mathbb{E}_M[\alpha]$ plot. Appendix~\ref{app:vqe_results} shows the results of VQEs with other gradient-based optimizers, there are no significant changes. Compared to VQEs, cQITE(P2A) lacks the ability to converge to the ground state, with most $p_{\text{gs}}$ values remaining at a low level. However, it benefits from a continuous growth of $p_{\text{gs}}$ before the energy starts to increase. As a result, the worst $p_{\text{gs}}$ observed for cQITE(P2A) ($0.018$) is significantly higher than that of the gradient-based methods trapped in local minima, such as VQE(iHVA) with a minimum $p_{\text{gs}}$ of $3.17\times10^{-6}$. Further gradient-based optimization provided by QIV indeed assists most of the cQITE-initialized states toward the ground states. However, in several instances, $p_{\text{gs}}$ becomes even worse when the initialization is not strong enough, as shown in Appendix~\ref{app:qiv}, underscoring the importance of continuous guidance from cQITE updates as adopted in AWQV. QITE(P2A) achieves the most comparable performance to the proposed AWQV, but AWQV requires only $\frac{1}{N}$ of the circuit depth of QITE(P2A), with $N = 50$ for these results.

\begin{table}[t]
    \caption{Optimization and simulation settings for MaxCut on unweighted graphs}
    \begin{center}
    \begin{tabular}{|c|c|c|}
    \hline
    \textbf{Method (Ansatz)} & \textbf{$\eta$/$\Delta\tau\ $} & \textbf{iteration/step ($N$)} \\
    \hline
    VQE(XQAOA) & 0.01 & 300 \\
    VQE(iHVA)  & 0.05 & 100 \\
    VQE(P2A)   & 0.05 & 50 \\
    cQITE(P1A) & 0.10 & 50 \\
    cQITE(P2A) & 0.05 & 50 \\
    cQITE(P2A-ZY) & 0.05 & 50 \\
    QITE(P2A)  & 0.05 & 50 \\
    QIV(P2A)   & 0.05 & 50 \\
    AWQV(P2A)  & 0.05 & 50 \\
    AWQV(P2A-ZY)  & 0.05 & 50 \\
    AWQV(P2A-XY)  & 0.05 & 50 \\
    \hline
    \end{tabular}
    \label{tab:train}
    \end{center}
\end{table}

To assess the contributions of different rotation types, we construct two simplified variants, AWQV(P2A-ZY) and AWQV(P2A-XY),  which remove the $R_{XY}$ and $R_{ZY}$ rotations from the original P2A circuit, respectively, thereby reducing the circuit depth by approximately half. As shown in Fig.~\ref{fig:n16reg3_xyzy}, However, unlike VQE(iHVA), AWQV(P2A-ZY) does not exhibit outliers with $p_{\text{gs}}$ values close to zero. As a result, when $M=10$, the expected approximation ratios for all instances approach 1. Thus, using P2A-ZY ansatz offers a favorable trade-off. Without $R_{ZY}$ rotations, as in AWQV(P2A-XY), $p_{\text{gs}}$ predominantly remain low, highlighting the critical contribution of $R_{ZY}$ rotations to the performance of AWQV.

We evaluate AWQV(P2A-ZY) against VQE(iHVA) and cQITE(P2A-ZY) on 16-vertex regular graphs with degrees ranging from 3 to 10. As shown in Fig.~\ref{fig:degree_size}(a), AWQV(P2A-ZY) performs consistently well across all degrees, with $p_{\text{gs}}$ values predominantly clustered around 0.5 and 1. Only a single exception is observed for a 3-regular graph instance, where $p_{\text{gs}} = 0.337$. In contrast, VQE(iHVA) displays instances with $p_{\text{gs}}$ values close to 0 appearing across almost all degree categories. cQITE(P2A-ZY) generally yields lower $p_{\text{gs}}$ values compared to AWQV(P2A-ZY) and VQE(iHVA). Notably, for several instances on 9-regular and 10-regular graphs, cQITE(P2A-ZY) reaches $p_{\text{gs}}$ levels comparable to those of VQE instances trapped in local minima, with detailed minimum values provided in Table~\ref{tab:mean_min_degree}.

We investigate the performance of AWQV on graphs of varying sizes by extending the comparison to 3-regular graphs with 12 to 24 vertices. As shown in Fig.~\ref{fig:degree_size}(b), the $p_{\text{gs}}$ clusters of both VQE(iHVA) and cQITE(P2A-ZY) gradually shift toward 0 as the graph size increases. In contrast, AWQV(P2A-ZY) maintains $p_{\text{gs}}$ values concentrated near 0.5 and 1, though an increased frequency of lower $p_{\text{gs}}$ outliers appears for larger instances. Table~\ref{tab:mean_min_size} summarizes the mean and minimum $p_{\text{gs}}$ values across different graph sizes for all three methods. For AWQV(P2A-ZY), the lowest observed $p_{\text{gs}}$ value is 0.109 for a 24-vertex graph. Since the expected number of measurements required to sample the optimal solution scales as $\frac{1}{p_{\text{gs}}}$, the expected sampling number is approximately 10 for 3-regular graphs up to 24 vertices.

\begin{table}[t]
    \caption{Mean and minimum $p_{\text{gs}}$ values on 16-vertex regular graphs with degrees 3 to 10}
    \begin{center}
    \begin{tabular}{|c|c|c|c|c|c|c|}
    \hline
    \textbf{Degree} & \multicolumn{2}{c|}{\textbf{AWQV(P2A-ZY)}} & \multicolumn{2}{c|}{\textbf{VQE(iHVA)}} & \multicolumn{2}{c|}{\textbf{cQITE(P2A-ZY)}} \\
    \cline{2-7}
     & Mean & Min & Mean & Min & Mean & Min \\
    \hline
    3  & 0.827 & 0.337 & 0.689 & 0.000 & 0.109 & 0.033 \\
    4  & 0.955 & 0.497 & 0.833 & 0.163 & 0.240 & 0.072 \\
    5  & 0.915 & 0.492 & 0.754 & 0.000 & 0.097 & 0.015 \\
    6  & 0.978 & 0.495 & 0.948 & 0.015 & 0.195 & 0.044 \\
    7  & 0.905 & 0.495 & 0.873 & 0.000 & 0.086 & 0.007 \\
    8  & 0.989 & 0.727 & 0.975 & 0.017 & 0.188 & 0.046 \\
    9  & 0.959 & 0.496 & 0.932 & 0.000 & 0.064 & 0.001 \\
    10 & 0.931 & 0.496 & 0.970 & 0.004 & 0.111 & 0.000 \\
    \hline
    \end{tabular}
    \label{tab:mean_min_degree}
    \end{center}
\end{table}

\begin{table}[t]
    \caption{Mean and minimum $p_{\text{gs}}$ values on 3-regular graphs with sizes from 12 to 24}
    \begin{center}
    \begin{tabular}{|c|c|c|c|c|c|c|}
    \hline
    \textbf{Size} & \multicolumn{2}{c|}{\textbf{AWQV(P2A-ZY)}} & \multicolumn{2}{c|}{\textbf{VQE(iHVA)}} & \multicolumn{2}{c|}{\textbf{cQITE(P2A-ZY)}} \\
    \cline{2-7}
     & Mean & Min & Mean & Min & Mean & Min \\
    \hline
    12  & 0.901 & 0.490 & 0.585 & 0.001 & 0.207 & 0.052 \\
    14  & 0.904 & 0.486 & 0.528 & 0.002 & 0.157 & 0.018 \\
    16  & 0.827 & 0.337 & 0.689 & 0.000 & 0.109 & 0.033 \\
    18  & 0.867 & 0.461 & 0.481 & 0.000 & 0.081 & 0.008 \\
    20  & 0.707 & 0.250 & 0.289 & 0.000 & 0.042 & 0.000 \\
    22  & 0.680 & 0.227 & 0.205 & 0.000 & 0.033 & 0.007 \\
    24  & 0.671 & 0.109 & 0.214 & 0.000 & 0.027 & 0.005 \\
    \hline
    \end{tabular}
    \label{tab:mean_min_size}
    \end{center}
\end{table}

\begin{figure*}[t]
    \centering
    \includegraphics[width=0.95\linewidth]{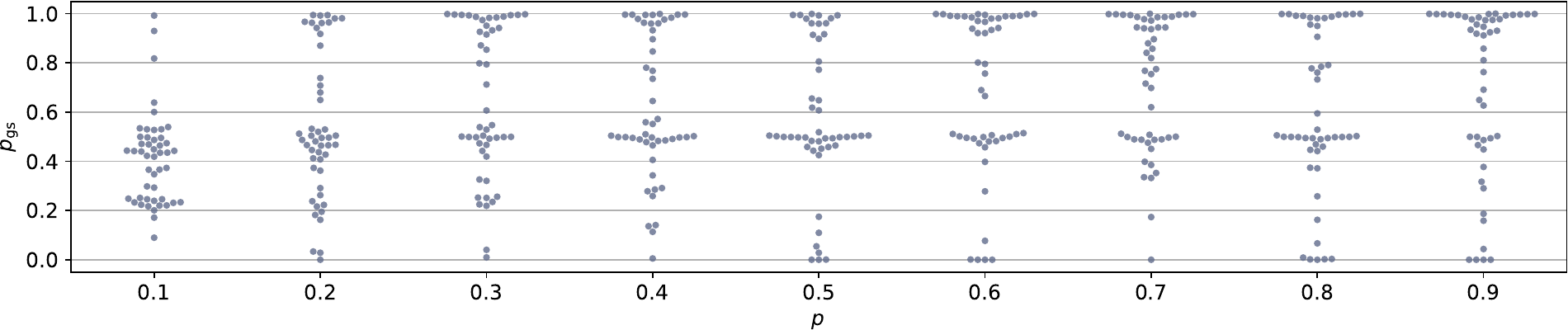}
    \caption{$p_{\text{gs}}$ values of states prepared by AWQV(P2A-ZY) for MaxCut on weighted Erdős–Rényi instances with varying edge probabilities $p$.}
    \label{fig:pgs_gnpw}
\end{figure*}

\begin{figure*}[t]
    \centering
    \includegraphics[width=0.95\linewidth]{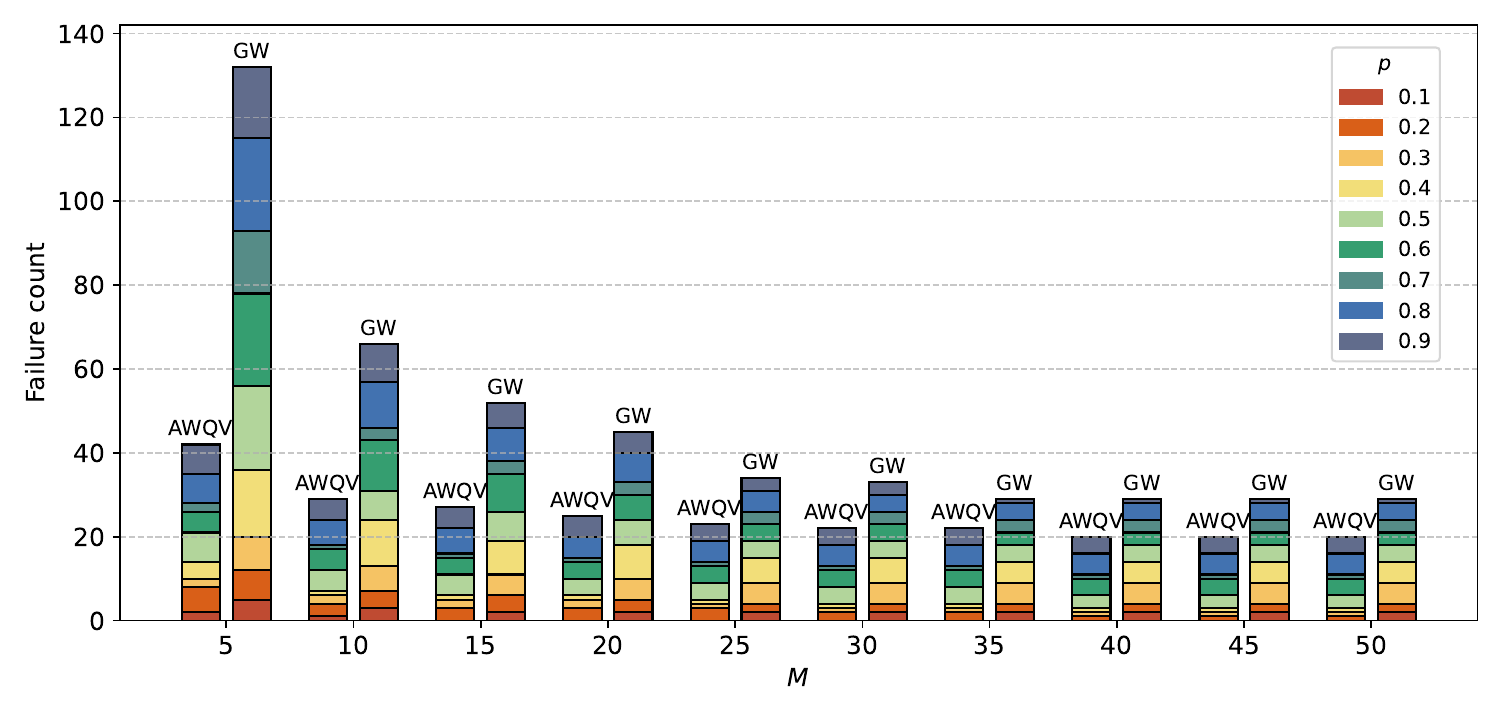}
    \caption{Failure counts versus sampling budget $M$ for AWQV and the GW algorithm on MaxCut problems. Results are shown for 432 weighted Erdős-Rényi instances with edge probabilities $p = 0.1, 0.2, \cdots, 0.9$. Different colors represent failures corresponding to graphs generated with different $p$.}
    \label{fig:failure_count}
\end{figure*}

\subsection{MaxCut on weighted graphs}
We next evaluate AWQV on weighted MaxCut problems using 16-vertex graphs randomly generated according to the Erdős-Rényi model, where each edge is included independently with probability $p\in\{0.1,0.2,\cdots,0.9\}$. For each $p$, we generate 48 random graphs with edge weights sampled from a standard random distribution. Here, we empirically set the hyperparameter $\mu$ in the weighting function $w$ to $0.8$.

As shown in Fig.~\ref{fig:pgs_gnpw}, AWQV exhibits reduced performance on weighted graphs, with several instances showing $p_{\text{gs}}$  values approaching 0, making it challenging to sample the optimal solution. We compare the failure rates in generating optimal solutions between AWQV and the GW algorithm. To ensure a fair comparison, we equate the number of random vectors used for hyperplane rounding in GW \cite{goemans1995improved} to the number of measurements $M$ performed on the prepared state in AWQV, as both methods fundamentally rely on random sampling mechanisms. We vary $M$ from $5$ to $50$ in the evaluation. For GW, we define failure as generating a non-optimal solution, while for AWQV, failure occurs when $M < \frac{1}{p_{\text{gs}}}$, i.e., the number of samples is insufficient to expect finding the optimal solution.


The failure counts shown in Fig.~\ref{fig:failure_count} suggest that AWQV has a significant advantage at smaller sampling budgets $M$. As $M$ increases, the gap between the two methods gradually narrows, particularly for graphs with $p>0.5$. Nevertheless, AWQV consistently maintains lower total failure counts across the tested range. GW stabilizes after $M = 35$, while AWQV stabilizes after $M = 40,$ ultimately retaining 29 and 20 failures, respectively.

\section{Conclusion}\label{sec:conclusion}
In this work, we proposed the AWQV algorithm, utilizing cQITE updates to guide the gradient-based optimization of VQE for CO problems. Numerical simulations on unweighted MaxCut instances demonstrate that AWQV consistently prepares states with high ground state probabilities, scaling up to 24-vertex graphs, while existing quantum approaches frequently converge to excited states. On weighted MaxCut instances, although the performance of AWQV reduces, it still achieves a lower failure ratio compared to the classical state-of-the-art, GW algorithm. Future work includes extending AWQV to a broader class of CO problems with a thorough scalability analysis across varying problem sizes, as well as conducting comparative studies with other promising variational approaches, such as VarQITE \cite{McArdle2019} and filtering variational quantum algorithms \cite{amaro2022filtering}.

\section*{Acknowledge}
This work was supported by JST SPRING, Grant Number JPMJSP2124.

\section*{Data Availability}
Data and code for the simulation are available at 
\href{https://github.com/NingyiXie/Adaptive-Weighted-QITE-VQE}{https://github.com/NingyiXie/Adaptive-Weighted-QITE-VQE}.

\appendix

\begin{figure}[t]
    \centering
    \includegraphics[width=0.95\linewidth]{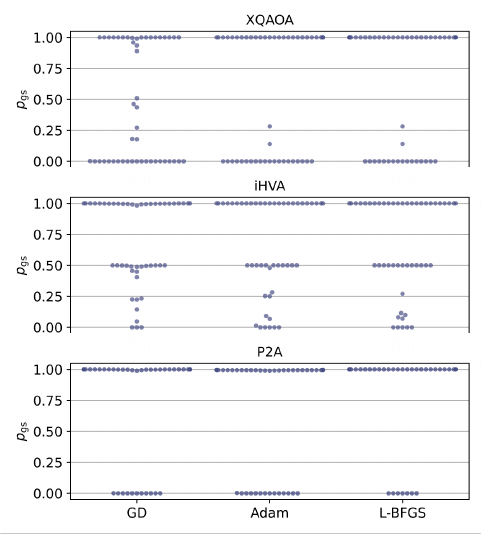}
    \caption{$p_{\text{gs}}$ values of states prepared by VQE(XQAOA), VQE(iHVA), and VQE(P2A) using GD, Adam \cite{kingm2015adam}, and L-BFGS \cite{Liu1989} optimizers for MaxCut on 48 randomly generated 16-vertex 3-regular graphs.}
    \label{fig:vqe_pgs}
\end{figure}

\begin{figure}[t]
    \centering
    \includegraphics[width=0.98\linewidth]{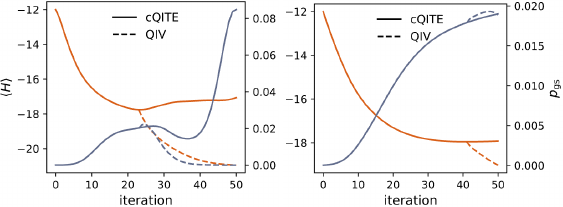}
    \caption{Progress of $\braket{H}$ (orange) and $p_{\text{gs}}$ (blue) during cQITE and subsequent QIV optimization for two representative instances of MaxCut on 16-vertex 3-regular graphs.}
    \label{fig:qiv_progress}
\end{figure}

\subsection{Proof of the equivalence of cQITE(P1A) to VQE with the GD optimizer}
\label{app:proof}

The Pauli set of P1A, also referred to as the linear ansatz in \cite{alam2023solving}, is defined by purely Pauli-$Y$ operators:
\begin{equation}
    \mathcal{P}_1^\ast = \left\{ Y_i \,\vert\, i \in \{1,2,\dots,n\} \right\},
\end{equation}
and the unitary operator is given by
\begin{equation}
    U(\boldsymbol{\theta}) = \prod_{j=1}^n e^{-i\frac{\theta_j}{2} Y_j}.
\end{equation}
Following \cite{alam2023solving}, the initial state is defined as
\begin{equation}
    \ket{\psi_0} = \bigotimes_{j=1}^{n} \ket{q_{0_j}},
\end{equation}
where each $\ket{q_{0_j}}$ belongs to $\{\ket{0}, \ket{1}, \ket{+}, \ket{-}\}$.  
Since there is no entanglement in the initial state and $e^{-i\frac{\theta}{2}Y_j}$ acts as a real rotation on a single qubit, preserving the real amplitudes, the intermediate state during cQITE evolution can be written as
\begin{equation}
    \ket{\psi} = \bigotimes_{j=1}^{n} \ket{q_j},
\end{equation}
with
\begin{equation}
    \ket{q_j} = c_{0,j}\ket{0} + c_{1,j}\ket{1},\quad c_{0,j},c_{1,j} \in \mathbb{R},\quad \forall j\in\{1,2,\dots,n\}.
\end{equation}

According to Eq.~(\ref{eqn:linear_system}), the matrix element $S_{jk}$ is
\begin{equation}
    \begin{aligned}
        S_{jk} &= \text{Re}\left( \braket{\psi| Y_j^\dagger Y_k | \psi} \right) \\
               &= \text{Re}\left( \braket{q_j| Y_j | q_j} \braket{q_k| Y_k | q_k} \right) \\
               &= \text{Re}\left(\begin{bmatrix} c_{0,j} & c_{1,j} \end{bmatrix} \begin{bmatrix} 0 & -i \\ i & 0 \end{bmatrix}\begin{bmatrix} c_{0,j} \\ c_{1,j} \end{bmatrix} \right.\\
                &\qquad\qquad\;\;\left.\begin{bmatrix} c_{0,k} & c_{1,k}\end{bmatrix} \begin{bmatrix} 0 & -i \\ i & 0 \end{bmatrix}\begin{bmatrix} c_{0,k} \\ c_{1,k} \end{bmatrix}\right) \\
               &= 0, \quad \text{if } j \neq k.
    \end{aligned}
\end{equation}
On the other hand, if $j = k$, then $S_{jk} = 1$. Therefore, $S$ is the identity matrix. The entries in the vector $\mathbf{b}$ can be expressed as
\begin{equation}
    \begin{aligned}
    b_j & = \text{Im}\left(\braket{\psi|Y_j^\dagger H|\psi}\right) \\
        & = -\left[\braket{\psi|\frac{i}{2}Y_jH|\psi}+\braket{\psi|\left(-\frac{i}{2}HY_j\right)|\psi}\right] \\
        & = -\left[\bra{\cdots q_{j-1}q_{0_j}q_{j+1}\cdots}\frac{i}{2}e^{i\frac{\theta_j}{2}Y_j}Y_jH\ket{\psi}+\right.\\
        & \quad\quad\;\; \left. \bra{\psi}H \left(-\frac{i}{2} Y_j e^{-i\frac{\theta_j}{2}Y_j}\right)\ket{\cdots q_{j-1}q_{0_j}q_{j+1}\cdots}\right]\\
        & = -\left[\bra{\cdots q_{j-1}q_{0_j}q_{j+1}\cdots}\frac{\partial e^{i\frac{\theta_j}{2}Y_j}}{\partial \theta_j}H\ket{\psi}+\right.\\
        & \quad\quad\;\; \left. \bra{\psi}H \frac{\partial e^{-i\frac{\theta_j}{2}Y_j}}{\partial \theta_j} \ket{\cdots q_{j-1}q_{0_j}q_{j+1}\cdots}\right]\\
        & = -\frac{\partial \braket{\psi|H|\psi}}{\partial \theta_j}.
    \end{aligned}
\end{equation}
Thus, we have
\begin{equation}
    \mathbf{b} = -\nabla_{\boldsymbol{\theta}} \braket{\psi| H | \psi}.
\end{equation}

According to Eq.~(\ref{eqn:cqite_update}), the update rule for $\boldsymbol{\theta}$ in cQITE is
\begin{equation}
    \begin{aligned}
        \boldsymbol{\theta} &\leftarrow \boldsymbol{\theta} + 2\Delta\tau S^{-1}\mathbf{b} \\
        &= \boldsymbol{\theta} - 2\Delta\tau \nabla_{\boldsymbol{\theta}} \braket{\psi|H|\psi},
    \end{aligned}
\end{equation}
which is exactly equivalent to applying GD with a learning rate $2\Delta\tau$. This concludes the proof.

\subsection{Results of VQE with different gradient-based optimizers}\label{app:vqe_results}
This section further examines Adam and L-BFGS optimizers for VQEs, as presented in Fig.~\ref{fig:vqe_pgs}. For VQE(P2A) and VQE(XQAOA), the use of L-BFGS slightly reduces the occurrence of instances with near-zero $p_{\text{gs}}$, whereas for VQE(iHVA), GD remains the most effective optimizer. Overall, the choice among GD, Adam, and L-BFGS does not lead to significant changes in the $p_{\text{gs}}$ distributions across the three ansatze.

\subsection{QIV under weak initialization}\label{app:qiv}
QIV switches to gradient-based optimization from the lowest-energy point found by cQITE. While the energy is further minimized, in two instances, the $p_{\text{gs}}$ values fail to improve or even decrease, as shown in Fig.~\ref{fig:qiv_progress}. In both cases, cQITE provided weak initializations, showing relatively high energy and low $p_{\text{gs}}$, indicating that the performance of QIV strongly depends on the quality of the initialization provided by cQITE.

\bibliographystyle{IEEEtran}
\bibliography{ref}

\begin{thebibliography}{10}
\providecommand{\url}[1]{#1}
\csname url@samestyle\endcsname
\providecommand{\newblock}{\relax}
\providecommand{\bibinfo}[2]{#2}
\providecommand{\BIBentrySTDinterwordspacing}{\spaceskip=0pt\relax}
\providecommand{\BIBentryALTinterwordstretchfactor}{4}
\providecommand{\BIBentryALTinterwordspacing}{\spaceskip=\fontdimen2\font plus
\BIBentryALTinterwordstretchfactor\fontdimen3\font minus \fontdimen4\font\relax}
\providecommand{\BIBforeignlanguage}[2]{{%
\expandafter\ifx\csname l@#1\endcsname\relax
\typeout{** WARNING: IEEEtran.bst: No hyphenation pattern has been}%
\typeout{** loaded for the language `#1'. Using the pattern for}%
\typeout{** the default language instead.}%
\else
\language=\csname l@#1\endcsname
\fi
#2}}
\providecommand{\BIBdecl}{\relax}
\BIBdecl

\bibitem{montanaro2016quantum}
A.~Montanaro, ``Quantum algorithms: an overview,'' \emph{npj Quantum Information}, vol.~2, no.~1, pp. 1--8, 2016.

\bibitem{grover1996fast}
L.~K. Grover, ``A fast quantum mechanical algorithm for database search,'' in \emph{Proceedings of the twenty-eighth annual ACM symposium on Theory of computing}, 1996, pp. 212--219.

\bibitem{shor1999polynomial}
P.~W. Shor, ``Polynomial-time algorithms for prime factorization and discrete logarithms on a quantum computer,'' \emph{SIAM review}, vol.~41, no.~2, pp. 303--332, 1999.

\bibitem{preskill2018quantum}
J.~Preskill, ``Quantum computing in the nisq era and beyond,'' \emph{Quantum}, vol.~2, p.~79, 2018.

\bibitem{peruzzo2014variational}
A.~Peruzzo, J.~McClean, P.~Shadbolt, M.-H. Yung, X.-Q. Zhou, P.~J. Love, A.~Aspuru-Guzik, and J.~L. O’brien, ``A variational eigenvalue solver on a photonic quantum processor,'' \emph{Nature communications}, vol.~5, no.~1, p. 4213, 2014.

\bibitem{farhi2014quantum}
E.~Farhi, J.~Goldstone, and S.~Gutmann, ``A quantum approximate optimization algorithm,'' \emph{arXiv preprint arXiv:1411.4028}, 2014.

\bibitem{farhi2000quantum}
E.~Farhi, J.~Goldstone, S.~Gutmann, and M.~Sipser, ``Quantum computation by adiabatic evolution,'' \emph{arXiv preprint quant-ph/0001106}, 2000.

\bibitem{hadfield2019quantum}
S.~Hadfield, Z.~Wang, B.~O’gorman, E.~G. Rieffel, D.~Venturelli, and R.~Biswas, ``From the quantum approximate optimization algorithm to a quantum alternating operator ansatz,'' \emph{Algorithms}, vol.~12, no.~2, p.~34, 2019.

\bibitem{wang2020xy}
Z.~Wang, N.~C. Rubin, J.~M. Dominy, and E.~G. Rieffel, ``Xy mixers: Analytical and numerical results for the quantum alternating operator ansatz,'' \emph{Physical Review A}, vol. 101, no.~1, p. 012320, 2020.

\bibitem{bartschi2020grover}
A.~B{\"a}rtschi and S.~Eidenbenz, ``Grover mixers for qaoa: Shifting complexity from mixer design to state preparation,'' in \emph{2020 IEEE International Conference on Quantum Computing and Engineering (QCE)}.\hskip 1em plus 0.5em minus 0.4em\relax IEEE, 2020, pp. 72--82.

\bibitem{herman2023constrained}
D.~Herman, R.~Shaydulin, Y.~Sun, S.~Chakrabarti, S.~Hu, P.~Minssen, A.~Rattew, R.~Yalovetzky, and M.~Pistoia, ``Constrained optimization via quantum zeno dynamics,'' \emph{Communications Physics}, vol.~6, no.~1, p. 219, 2023.

\bibitem{xie2024feasibility}
N.~Xie, X.~Lee, D.~Cai, Y.~Saito, N.~Asai, and H.~C. Lau, ``A feasibility-preserved quantum approximate solver for the capacitated vehicle routing problem,'' \emph{Quantum Information Processing}, vol.~23, no.~8, p. 291, 2024.

\bibitem{sack2024large}
S.~H. Sack and D.~J. Egger, ``Large-scale quantum approximate optimization on nonplanar graphs with machine learning noise mitigation,'' \emph{Physical Review Research}, vol.~6, no.~1, p. 013223, 2024.

\bibitem{maciejewski2024improving}
F.~B. Maciejewski, J.~Biamonte, S.~Hadfield, and D.~Venturelli, ``Improving quantum approximate optimization by noise-directed adaptive remapping,'' \emph{arXiv preprint arXiv:2404.01412}, 2024.

\bibitem{tan2021qubit}
B.~Tan, M.-A. Lemonde, S.~Thanasilp, J.~Tangpanitanon, and D.~G. Angelakis, ``Qubit-efficient encoding schemes for binary optimisation problems,'' \emph{Quantum}, vol.~5, p. 454, 2021.

\bibitem{Glos2022}
A.~Glos, A.~Krawiec, and Z.~Zimborás, ``Space-efficient binary optimization for variational quantum computing,'' \emph{npj Quantum Information}, vol.~8, no.~1, p.~39, 2022.

\bibitem{herrman2022multi}
R.~Herrman, P.~C. Lotshaw, J.~Ostrowski, T.~S. Humble, and G.~Siopsis, ``Multi-angle quantum approximate optimization algorithm,'' \emph{Scientific Reports}, vol.~12, no.~1, p. 6781, 2022.

\bibitem{vijendran2024expressive}
V.~Vijendran, A.~Das, D.~E. Koh, S.~M. Assad, and P.~K. Lam, ``An expressive ansatz for low-depth quantum approximate optimisation,'' \emph{Quantum Science and Technology}, vol.~9, no.~2, p. 025010, 2024.

\bibitem{wang2025imaginary}
X.~Wang, Y.~Chai, X.~Feng, Y.~Guo, K.~Jansen, and C.~T{\"u}ys{\"u}z, ``Imaginary hamiltonian variational ansatz for combinatorial optimization problems,'' \emph{Physical Review A}, vol. 111, no.~3, p. 032612, 2025.

\bibitem{goemans1995improved}
M.~X. Goemans and D.~P. Williamson, ``Improved approximation algorithms for maximum cut and satisfiability problems using semidefinite programming,'' \emph{Journal of the ACM (JACM)}, vol.~42, no.~6, pp. 1115--1145, 1995.

\bibitem{mcclean2018barren}
J.~R. McClean, S.~Boixo, V.~N. Smelyanskiy, R.~Babbush, and H.~Neven, ``Barren plateaus in quantum neural network training landscapes,'' \emph{Nature communications}, vol.~9, no.~1, p. 4812, 2018.

\bibitem{lee2021parameters}
X.~Lee, Y.~Saito, D.~Cai, and N.~Asai, ``Parameters fixing strategy for quantum approximate optimization algorithm,'' in \emph{2021 IEEE international conference on quantum computing and engineering (QCE)}.\hskip 1em plus 0.5em minus 0.4em\relax IEEE, 2021, pp. 10--16.

\bibitem{blekos2024review}
K.~Blekos, D.~Brand, A.~Ceschini, C.-H. Chou, R.-H. Li, K.~Pandya, and A.~Summer, ``A review on quantum approximate optimization algorithm and its variants,'' \emph{Physics Reports}, vol. 1068, pp. 1--66, 2024.

\bibitem{zhou2020quantum}
L.~Zhou, S.-T. Wang, S.~Choi, H.~Pichler, and M.~D. Lukin, ``Quantum approximate optimization algorithm: Performance, mechanism, and implementation on near-term devices,'' \emph{Physical Review X}, vol.~10, no.~2, p. 021067, 2020.

\bibitem{alam2020accelerating}
M.~Alam, A.~Ash-Saki, and S.~Ghosh, ``Accelerating quantum approximate optimization algorithm using machine learning,'' in \emph{2020 Design, Automation \& Test in Europe Conference \& Exhibition (DATE)}.\hskip 1em plus 0.5em minus 0.4em\relax IEEE, 2020, pp. 686--689.

\bibitem{moussa2022unsupervised}
C.~Moussa, H.~Wang, T.~B{\"a}ck, and V.~Dunjko, ``Unsupervised strategies for identifying optimal parameters in quantum approximate optimization algorithm,'' \emph{EPJ Quantum Technology}, vol.~9, no.~1, p.~11, 2022.

\bibitem{xie2023quantum}
N.~Xie, X.~Lee, D.~Cai, Y.~Saito, and N.~Asai, ``Quantum approximate optimization algorithm parameter prediction using a convolutional neural network,'' in \emph{Journal of Physics: Conference Series}, vol. 2595, no.~1.\hskip 1em plus 0.5em minus 0.4em\relax IOP Publishing, 2023, p. 012001.

\bibitem{lee2023depth}
X.~Lee, N.~Xie, D.~Cai, Y.~Saito, and N.~Asai, ``A depth-progressive initialization strategy for quantum approximate optimization algorithm,'' \emph{Mathematics}, vol.~11, no.~9, p. 2176, 2023.

\bibitem{motta2020determining}
M.~Motta, C.~Sun, A.~T. Tan, M.~J. O’Rourke, E.~Ye, A.~J. Minnich, F.~G. Brandao, and G.~K.-L. Chan, ``Determining eigenstates and thermal states on a quantum computer using quantum imaginary time evolution,'' \emph{Nature Physics}, vol.~16, no.~2, pp. 205--210, 2020.

\bibitem{McArdle2019}
S.~McArdle, T.~Jones, S.~Endo, Y.~Li, S.~C. Benjamin, and X.~Yuan, ``Variational ansatz-based quantum simulation of imaginary time evolution,'' \emph{npj Quantum Information}, vol.~5, no.~1, p.~75, 2019.

\bibitem{mclachlan1964variational}
A.~D. McLachlan, ``A variational solution of the time-dependent schrodinger equation,'' \emph{Molecular Physics}, vol.~8, no.~1, pp. 39--44, 1964.

\bibitem{broeckhove1988equivalence}
J.~Broeckhove, L.~Lathouwers, E.~Kesteloot, and P.~Van~Leuven, ``On the equivalence of time-dependent variational principles,'' \emph{Chemical physics letters}, vol. 149, no. 5-6, pp. 547--550, 1988.

\bibitem{stokes2020quantum}
J.~Stokes, J.~Izaac, N.~Killoran, and G.~Carleo, ``Quantum natural gradient,'' \emph{Quantum}, vol.~4, p. 269, 2020.

\bibitem{nishi2021implementation}
H.~Nishi, T.~Kosugi, and Y.-i. Matsushita, ``Implementation of quantum imaginary-time evolution method on nisq devices by introducing nonlocal approximation,'' \emph{npj Quantum Information}, vol.~7, no.~1, p.~85, 2021.

\bibitem{gomes2020efficient}
N.~Gomes, F.~Zhang, N.~F. Berthusen, C.-Z. Wang, K.-M. Ho, P.~P. Orth, and Y.~Yao, ``Efficient step-merged quantum imaginary time evolution algorithm for quantum chemistry,'' \emph{Journal of Chemical Theory and Computation}, vol.~16, no.~10, pp. 6256--6266, 2020.

\bibitem{zhang2022variational}
Y.~Zhang, L.~Cincio, C.~F. Negre, P.~Czarnik, P.~J. Coles, P.~M. Anisimov, S.~M. Mniszewski, S.~Tretiak, and P.~A. Dub, ``Variational quantum eigensolver with reduced circuit complexity,'' \emph{npj Quantum Information}, vol.~8, no.~1, p.~96, 2022.

\bibitem{yan2024light}
X.~Yan, X.~Lee, N.~Xie, Y.~Saito, L.~Kurosawa, N.~Asai, D.~Cai, and H.~C. Lau, ``Light cone cancellation for variational quantum eigensolver ansatz,'' \emph{arXiv preprint arXiv:2404.19497}, 2024.

\bibitem{ruder2016overview}
S.~Ruder, ``An overview of gradient descent optimization algorithms,'' \emph{arXiv preprint arXiv:1609.04747}, 2016.

\bibitem{mitarai2018quantum}
K.~Mitarai, M.~Negoro, M.~Kitagawa, and K.~Fujii, ``Quantum circuit learning,'' \emph{Physical Review A}, vol.~98, no.~3, p. 032309, 2018.

\bibitem{schuld2019evaluating}
M.~Schuld, V.~Bergholm, C.~Gogolin, J.~Izaac, and N.~Killoran, ``Evaluating analytic gradients on quantum hardware,'' \emph{Physical Review A}, vol.~99, no.~3, p. 032331, 2019.

\bibitem{alam2023solving}
R.~Alam, G.~Siopsis, R.~Herrman, J.~Ostrowski, P.~C. Lotshaw, and T.~S. Humble, ``Solving maxcut with quantum imaginary time evolution,'' \emph{Quantum Information Processing}, vol.~22, no.~7, p. 281, 2023.

\bibitem{bauer2024combinatorial}
N.~M. Bauer, R.~Alam, G.~Siopsis, and J.~Ostrowski, ``Combinatorial optimization with quantum imaginary time evolution,'' \emph{Physical Review A}, vol. 109, no.~5, p. 052430, 2024.

\bibitem{rasmussen2008round}
R.~V. Rasmussen and M.~A. Trick, ``Round robin scheduling--a survey,'' \emph{European Journal of Operational Research}, vol. 188, no.~3, pp. 617--636, 2008.

\bibitem{Zhang2023tensorcircuit}
S.-X. Zhang, J.~Allcock, Z.-Q. Wan, S.~Liu, J.~Sun, H.~Yu, X.-H. Yang, J.~Qiu, Z.~Ye, Y.-Q. Chen, C.-K. Lee, Y.-C. Zheng, S.-K. Jian, H.~Yao, C.-Y. Hsieh, and S.~Zhang, ``Tensor{C}ircuit: a {Q}uantum {S}oftware {F}ramework for the {NISQ} {E}ra,'' \emph{{Quantum}}, vol.~7, p. 912, Feb. 2023.

\bibitem{amaro2022filtering}
D.~Amaro, C.~Modica, M.~Rosenkranz, M.~Fiorentini, M.~Benedetti, and M.~Lubasch, ``Filtering variational quantum algorithms for combinatorial optimization,'' \emph{Quantum Science and Technology}, vol.~7, no.~1, p. 015021, 2022.

\bibitem{kingm2015adam}
D.~P. Kingma and J.~Ba, ``Adam: A method for stochastic optimization,'' in \emph{International Conference on Learning Representations (ICLR)}, 2015.

\bibitem{Liu1989}
D.~C. Liu and J.~Nocedal, ``On the limited memory bfgs method for large scale optimization,'' \emph{Mathematical Programming}, vol.~45, no.~1, pp. 503--528, 1989.

\end{thebibliography}

\end{document}